\documentclass[12pt]{article}

\usepackage{graphicx}
\usepackage{fullpage}
\usepackage{amsmath}
\usepackage{amsthm}

\newtheorem{theorem}{Theorem}
\newtheorem{cor}[theorem]{Corollary}
\newtheorem{lemma}[theorem]{Lemma}
\theoremstyle{definition}

\newtheorem{rem}[theorem]{Remark}

\begin{document}
\author{Anthony J. Guttmann$^\dagger$ \and  Robert Parviainen$^\ddagger$ \and Andrew Rechnitzer$^\ast$\\
  {\small ARC Centre of Excellence for Mathematics and Statistics of Complex Systems and }\\
  {\small Department of Mathematics and Statistics,} \\
  {\small The University of Melbourne, Parkville Victoria 3010, Australia.}\\
  {\small $\dagger$ \texttt{tonyg@ms.unimelb.edu.au} $\ddagger$ \texttt{robertp@ms.unimelb.edu.au} $\ast$ 
    \texttt{a.rechnitzer@ms.unimelb.edu.au}} \\
}
%\date{\today}
\title{Self-avoiding walks and trails on the $3.12^2$ lattice}
\maketitle

\begin{abstract}
  We find the generating function of self-avoiding walks and trails on
  a semi-regular lattice called the $3.12^2$ lattice in terms of the
  generating functions of simple graphs, such as self-avoiding walks,
  polygons and tadpole graphs on the hexagonal lattice. Since the
  growth constant for these graphs is known on the hexagonal lattice
  we can find the growth constant for both walks and trails on the
  $3.12^2$ lattice. A result of Watson \cite{Watson1970} then allows
  us to find the generating function and growth constant of
  neighbour-avoiding walks on the covering lattice of the $3.12^2$
  lattice which is tetra-valent. A mapping into walks on the covering lattice
allows us to obtain improved bounds on the growth constant for a range of lattices.
\end{abstract}

 \section{Introduction}
The enumeration of self-avoiding walks and polygons (SAWs and SAPs
respectively) are long-standing combinatorial problems with many
connections to statistical mechanics and theoretical chemistry. The
vast majority of results for these models on two-dimensional lattices concern
SAWs and SAPs embedded in one of the three regular planar lattices,
the triangular, the square or the hexagonal lattice. 

More recently, attention has turned to other two-dimensional lattices,
including semi-regular lattices (also known as Archimedian or
homogeneous tilings \cite{Jensen1998}) quasi-periodic lattices
\cite{RRG03} and non-Euclidean hyperbolic lattices \cite{SG96}.

\begin{figure}[h!]
  \centering
  \includegraphics[scale=0.4]{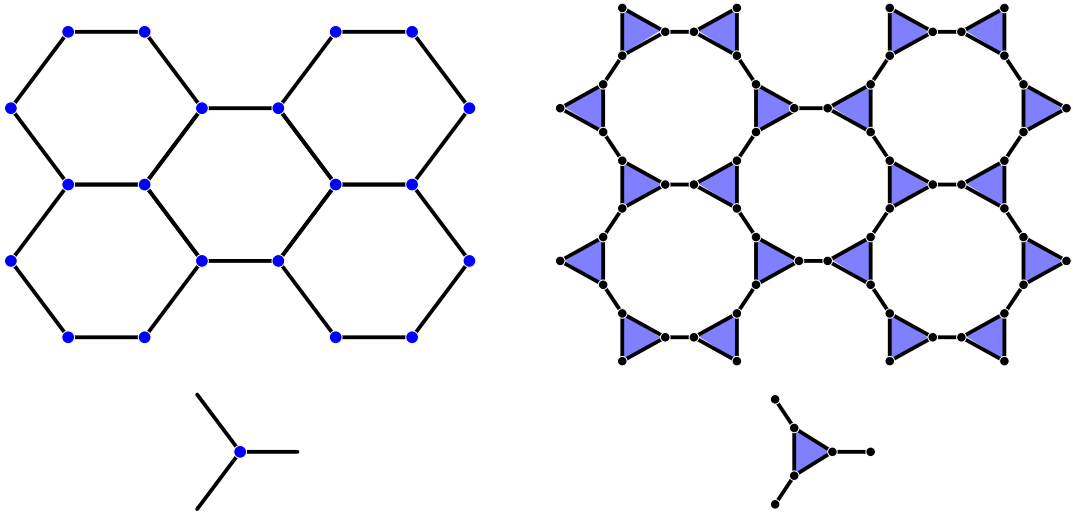}
  \caption{The $3.12^2$ lattice (right) can be constructed from the
    hexagonal lattice (left) by replacing each vertex with a triangle
    as shown.}
  \label{fig 312 lattice}
\end{figure}

In an earlier paper Jensen and Guttmann \cite{Jensen1998} studied the
number of SAPs on the semi-regular $3.12^2$ lattice (see
Figure~\ref{fig 312 lattice}). The $3.12^2$ lattice can be constructed
from the hexagonal lattice by replacing each vertex with a triangle as
shown in Figure~\ref{fig 312 lattice}.

While the number of SAWs on this lattice was also discussed in \cite{Jensen1998}
there is an error in their result due to the effects of the endpoints of the walks
 --- this was noted by
Alm and Parviainen \cite{Alm2004}. In this paper we correct this error
and correctly compute the number of SAWs on the $3.12^2$ lattice in terms of the
number of SAWs (and related graphs) on the hexagonal lattice.
Similar reasoning then allows us to extend this result to the number of
self-avoiding trails (SATs).

In Section~\ref{sec 312 hex} we obtain the generating functions for SAWs
and SATs on the $3.12^2$ lattice. This also gives the growth
constant for SAWs on this lattice. In Section~\ref{sec covering} we give a
general bijection between SAWs on a given lattice and SAWs with no
nearest-neighbour contacts on the \emph{covering lattice}.
  From this
bijection we find the growth constant of neighbour-avoiding SAW on this new lattice
which is tetra-valent.

\section{Mapping between the $3.12^2$ lattice and the hexagonal lattice}
\label{sec 312 hex}
\begin{figure}[h!]
  \centering
  \includegraphics[scale=0.35]{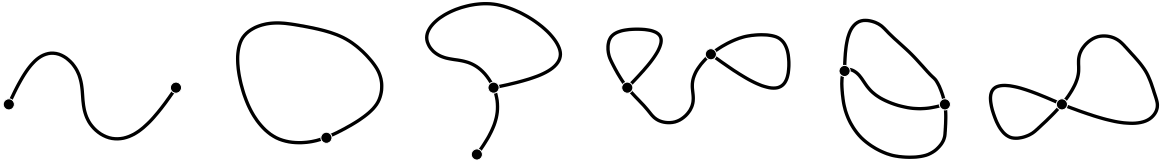}
  \caption{(From left to right) Walks, returns, tadpoles, dumbbells,
    theta graphs and figure-eights.}
  \label{fig hex objects}
\end{figure}

In this section we calculate the generating functions of SAWs and SATs
on the $3.12^2$ lattice counted by the number of edges. Let these
generating functions be $S_w(z)$ and $S_t(z)$ respectively. The first
few terms of these generating functions are given in Table~\ref{tab
  312 series}; they were computed using a recursive backtracking
algorithm. Not surprisingly, we have been unable to find closed-form
expressions for these generating functions. However, by constructing a
mapping between the $3.12^2$ lattice and the hexagonal lattice we can
express $S_w(z)$ and $S_t(z)$ in terms of the generating functions of
graphs (counted by the number of edges) embedded in the hexagonal
lattice. In particular, we require the following generating functions:
\begin{itemize}
\item Let $W(z)$ be the generating function of self-avoiding walks, without
the constant term. That is to say, the first term is O$(z)$.
\item Let $R(z)$ be the generating function of returns (or SAPs with
  a single marked vertex),
\item Let $T(z)$ be the generating function of tadpoles, 
\item and finally let $\Theta(z),$  $D(z)$ and $E(z)$  be the generating
  functions of theta graphs, dumbbells and figure-eights
     respectively.
(Note that figure-eights have a vertex of degree 4, and so are not embeddable
in the $3.12^2$ or hexagonal lattices.)
\end{itemize}
Pictures of these graphs are given in Figure~\ref{fig hex objects},
and the first few coefficients of the corresponding generating
functions are given in Table~\ref{tab hex series} .

\begin{figure}[h!]
  \centering
  \includegraphics[scale=0.4]{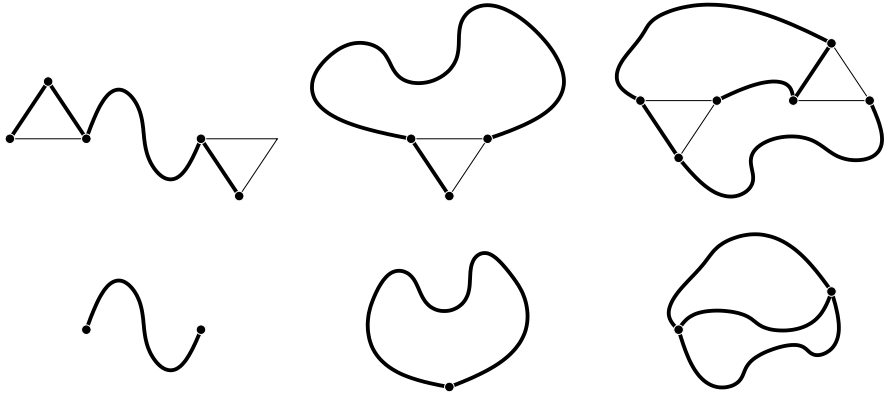}
  \caption{Examples of mapping walks from the $3.12^2$ lattice (top) to
    graphs on the hexagonal lattice (bottom). Depending on the
    positions of the endpoints of the original SAW it may map to
    (among other possibilities) a walk, a return or a theta graph. }
  \label{fig map eg}
\end{figure}

The main result of the paper 
 is the following theorem:
\begin{theorem}
  \label{thm hex 312 mapping}
  The generating function of self-avoiding walks on the $3.12^2$
  lattice is given by
  \begin{displaymath}
    S_w(z) = e + \frac{e^2}{3v} W(zv) + \frac{m}{3v} R(zv) 
    + \frac{2de}{3v^2}T(zv) 
    + \frac{2d^2}{3v^3}D(zv) + \frac{d^2}{v^3}\Theta(zv)
  \end{displaymath}
  where $v=z+z^2, e = 1+2z+2z^2, m = 1+2z$ and $d=2z$.

  Similarly, the generating function of self-avoiding trails on the
  $3.12^2$ lattice is given by
  \begin{displaymath}
    S_t(z) = e + \frac{e^2}{3v} W(zv) + \frac{m}{3v} R(zv) 
    + \frac{2de}{3v^2}T(zv) 
    + \frac{2d^2}{3v^3} D(zv) + \frac{d^2}{v^3}\Theta(zv)
  \end{displaymath}
  where $v = z+z^2, e = 1+2z+2z^2+2z^3, m = 1+4z+7z^2+6z^3$ and $d=2z+6z^2$.

\end{theorem}
Incidentally, our enumeration to order $z^{35}$  is fortunately consistent with the theorem.
\proof Let $\varphi$ be a self-avoiding walk on the $3.12^2$ lattice.
We map $\varphi$ to an object $\psi$ on the hexagonal lattice by
contracting each triangular face to a vertex while keeping track of the
location of the endpoints. Depending on the locations of
the endpoints of $\varphi$, $\psi$ can be a walk, a return, a tadpole,
a dumbbell or a theta graph (see Figure~\ref{fig map eg} for examples).
In order to determine which family of graphs $\psi$ belongs to, we
need to consider the possible bond configurations of the triangular
faces incident upon the original walk $\varphi$ (see Figure~\ref{fig
  map vertices}). In particular if a triangular face is connected to
\begin{itemize}
\item one edge of $\varphi$, then it is an $e$-triangle
\item two edges of $\varphi$, (which connect outside the triangle) then it is an $m$-triangle
\item two edges of $\varphi$, (that connect via the triangle) then it a $v$-triangle
\item three edges of $\varphi$, then it is a $d$-triangle.
\end{itemize}
Since $\varphi$ contains only two endpoints, the graph $\psi$ must
belong to one of only six sets; if $\varphi$ contains:
\begin{itemize}
\item a single $e$ triangle, then $\psi$ is a single vertex
\item two $e$ triangles, then $\psi$ is a walk
\item a single $m$ triangle, then $\psi$ is a return
\item one $e$ and one $d$ triangle, then $\psi$ is a tadpole
\item two $d$ triangles, then $\psi$ is either a dumbbell or a theta graph.
\end{itemize}

\begin{figure}[h!]
  \centering
  \includegraphics[scale=0.35]{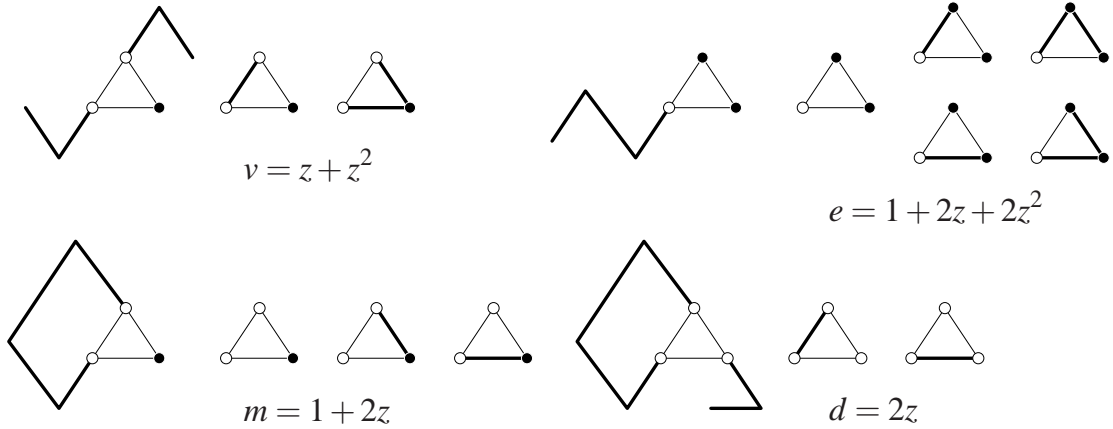}
  \caption{The different types of triangles for the SAW mapping and
    the possible bond configurations within them.}
  \label{fig map vertices}
\end{figure}

We define a reverse mapping $\psi \mapsto \varphi$ by expanding each
vertex on the hexagonal lattice to form a triangle. This is a
one-to-many mapping.
\begin{itemize}
\item A single vertex becomes an $e$-triangle.
\item A walk of $n$ edges contains $2$ vertices that map to
  $e$-triangles, and $n-1$ vertices that map to $v$-triangles.
\item A return of $n$ edges contains $1$ marked vertex that becomes an
  $m$-triangle and $n-1$ vertices that become $v$-triangles.
\item A tadpole of $n$-edges contains a single degree 1 vertex which becomes an
  $e$-triangle and a single degree 3 vertex that becomes a
  $d$-triangle. The remaining $n-2$ vertices become
  $v$-triangles.  
\item A theta graph or dumbbell contains two vertices of degree 3 which
  become $d$-triangles and the remaining $n-3$ vertices become $v$-triangles.
\end{itemize}
These possibilities give rise to each of the terms in the equations in
Theorem~\ref{thm hex 312 mapping}. The contributions of the different
triangles are given in Figure~\ref{fig map vertices} --- for example
the contribution from the case when $\psi$ is a walk gives rise to
$\frac{e^2}{3v} W(zv) = \frac{(1+2z+2z^2)^2}{3(z+z^2)}
W(z^2+z^3)$. The
prefactors arise from the number of orientations and symmetries of the
graphs.

\begin{figure}[h!]
  \centering
  \includegraphics[scale=0.35]{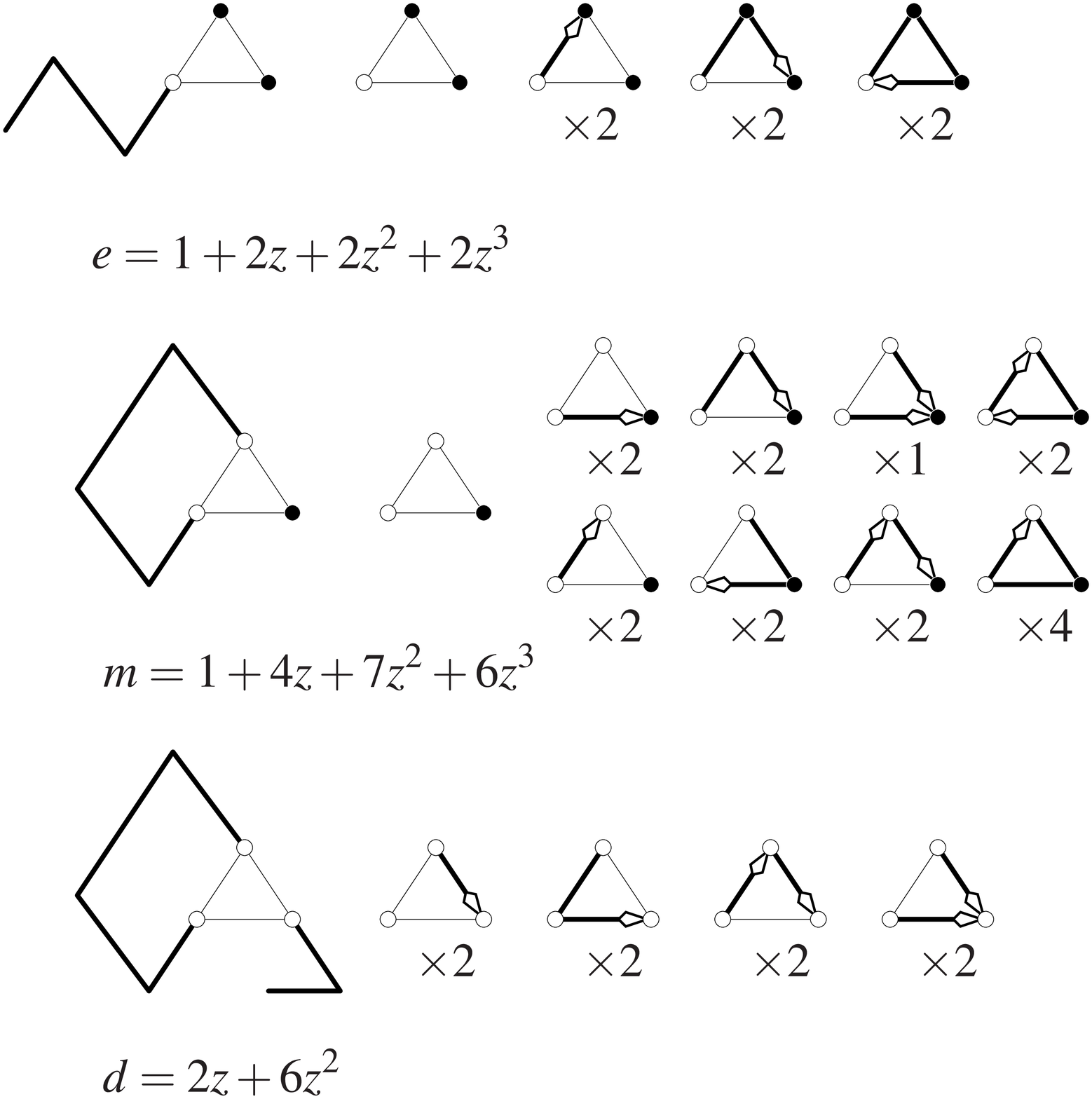}
  \caption{The different types of triangles for the SAT mapping and
    the possible bond configurations within them.}
  \label{fig map trail vertices}
\end{figure}

The proof for trails is very similar except that the endpoint
considerations are a little more complicated. Again a trail $\varphi$
maps to a graph $\psi$ which is either a single vertex, a walk, a
return, a tadpole, a dumbbell or a theta graph. The contributions of $e$-,
$m$- and $d$-triangles are slightly different. See Figure~\ref{fig
  map trail vertices}.
\qed

\vspace{4ex}

We can improve these results by using Sykes's Counting Theorem
\cite{S61}, which relates the SAW generating function to the
generating functions of the other families of graphs listed above. This
allows us to eliminate some generating functions from the above
results. In particular we are able to eliminate either the
contribution of tadpoles, or returns, or both theta graphs and dumbbells.
We elect to remove the last two. 
% For a lattice of co-ordination
%number $q$ we have \cite{S61}
\begin{theorem}[Sykes Counting Theorem \cite{S61}]
  \label{thm tadpole1}
  The generating function of tadpoles on a lattice of co-ordination
  number $q$ is given by:
  \begin{displaymath}
%    T(z) =  \frac{z(q-1)(z(q-1) - 1)}{2}W(z) + \frac{z^2q(q-1)}{2}-\frac{z}{2}R(z)-
%4E(z) - 6\Theta(z)-4D(z).
    T(z) =  \frac{z^2(q-1)^2 - z(q-1)}{2}W(z) + \frac{z^2q(q-1)}{2}-\frac{z}{2}R(z)-
4E(z) - 6\Theta(z)-4D(z).
  \end{displaymath}
  For the $3.12^2$ lattice, $q=3$ and the result becomes:
  \begin{displaymath}
    T(z) =  W(z)(2z^2 - z) + 3z^2-\frac{z}{2}R(z)-
    6\Theta(z)-4D(z).
  \end{displaymath}
\end{theorem}
Eliminating the term involving the dumbbell and theta-graph generating
functions from Theorem~\ref{thm hex 312 mapping} gives:
\begin{theorem}
  \label{thm no theta db}
  The generating function of self-avoiding walks on the $3.12^2$
  lattice is given by:
  \begin{displaymath}
    S_w(z) = P_0(z) + P_1(z)W(zv) + P_2(z) R(zv) 
    + P_3(z)T(zv) 
  \end{displaymath}
  where $v=z+z^2$ and the $P_i(z)$ are given by\\
  \begin{displaymath}
  \begin{array}{ll}
    P_0(z)=\frac{1+3z+4z^2+4z^3}{1+z}, &
    P_1(z)=\frac{1+5z+10z^2+16z^3+16z^4+8z^5}{3z(1+z)^2}, \\
    P_2(z)=\frac{1+3z+z^2}{3z(1+z)^2} \mbox{ and } &
    P_3(z)=\frac{2(1+6z+8z^2+4z^3)}{3z(1+z)^3}.
  \end{array}
\end{displaymath}
  Similarly the generating function of self-avoiding trails on the
  $3.12^2$ lattice is given by
  \begin{displaymath}
    S_t(z) = Q_0(z) + Q_1(z)W(zv) + Q_2(z) R(zv) 
    + Q_3(z)T(zv) 
  \end{displaymath}
  where $v=z+z^2$ and the $Q_i(z)$ are given by
  \begin{displaymath}
    \begin{array}{ll}
      Q_0(z)=\frac{1+3z+4z^2+6z^3+14z^4+18z^5}{1+z}, &
      Q_1(z)=\frac{1+5z+10z^2+8z^3+10z^4+48z^5+72z^6+40z^7}{3z(1+z)^2}, \\
      Q_2(z)=\frac{1+5z+10z^2+7z^3-3z^4}{3z(1+z)^2} \mbox{ and } &
      Q_3(z)=\frac{2(1+3z)(1+3z+8z^2+8z^3+4z^4)}{3z(1+z)^3}.
    \end{array}
  \end{displaymath}
\end{theorem}
\proof This follows by eliminating the contributions of the theta graph and
dumbbell generating functions between Theorems~\ref{thm hex 312
  mapping}~and~\ref{thm tadpole1}. \qed

\vspace{4ex}

These generating functions then allow us to obtain a number of results
on the growth constants for various models on various lattices.
The growth constant of self-avoiding walks is defined as the limit
\begin{equation}
  \lim_{n \to \infty} c_n^{1/n} = \mu.
\end{equation}
where $c_n$ is the number of SAWs of length $n$. Hammersley
\cite{Hammersley1957} first proved that this limit exists and is
finite. The value of $\mu$ is not known \emph{rigorously} on any
two-dimensional lattice, however an argument due to Nienhuis
\cite{Nienhuis1982} implies that on the hexagonal lattice
\begin{equation}
  \lim_{n \to \infty} c_n^{1/n} = \mu = \sqrt{2+\sqrt{2}}.
\end{equation}
While this argument is \emph{not entirely rigorous}, the result is
supported by strong theoretical and numerical evidence
\cite{Nienhuis1982, EG89} and is widely accepted. More recently Jensen
\cite{J04} and Alm and Parviainen \cite{Alm2004} give rigorous bounds
on the value of $\mu$.
\begin{equation}
  1.841925 < \mu \approx 1.8477590\dots < 1.868832
\end{equation}
Using the believed exact value of $\mu$ and Theorem~\ref{thm hex 312
  mapping} or~\ref{thm no theta db}, we are able to find the growth
constant of walks and trails on the $3.12^2$ lattice.
\begin{cor}\label{cor 31212}
  The growth constant for self-avoiding walks and self-avoiding
  trails on the $3.12^2$ lattice is the largest positive real solution of
  \begin{displaymath}
    \lambda^{12}-4\lambda^8-8\lambda^7-4\lambda^6+2\lambda^4+8\lambda^3+12\lambda^2+8\lambda+2 = 0
  \end{displaymath}
  which is
  \begin{displaymath}
    \lambda = 1.711041297\dots
  \end{displaymath}
  Note that this relies on Nienhuis' result that $\mu =
  \sqrt{2+\sqrt{2}}$ on the hexagonal lattice.

  We also note that this agrees with the rigorous bounds given in
  \cite{J04} and \cite{Alm2004} of
  \begin{displaymath}
    1.708553 < \lambda < 1.719254
  \end{displaymath}
\end{cor}
\proof We find the growth constant of SAWs and SATs on the
$3.12^2$ lattice by examining the radius of convergence of their
generating functions. In particular if the radius of convergence is
$\rho$, then the growth constant is $1/\rho$.

Theorem~\ref{thm hex 312 mapping} implies that the singularities of
$S_w(z)$ and $S_t(z)$ arise from the singularities of $W(z^2+z^3),
R(z^2+z^3), T(z^2+z^3), D(z^2+z^3)$, and $\Theta(z^2+z^3)$ and from
the factors of $\frac{1}{z(1+z)}$.

There is no singularity at $z=0$, since the factors are cancelled by
factors of $z$ in the hexagonal lattice series, and the factors of
$1/(1+z)$ give singularities at $z=-1$.

It was shown by Guttmann and Whittington \cite{Guttmann1978} that
returns, tadpoles, dumbbells and theta graphs all have the same
growth constant as self-avoiding walks --- their results are for
the square lattice, but apply, {\em mutatis mutandis}, to the hexagonal
lattice. 

Hence the generating functions $W(x), R(x), T(x), D(x)$, and
$\Theta(x)$ all have a dominant singularity at $x=(
2+\sqrt{2})^{-1/2}$ and so the generating functions $S_w(z)$ and
$S_t(z)$ have dominant singularities at $z=z_c$ satisfying $z_c^2+z_c^3 =
(2+\sqrt{2})^{-1/2}$. Some manipulation of this equation implies that
the growth constant must satisfy the equation given in the statement
of the corollary.

\qed

We note that this is the same as the growth constant for self-avoiding
polygons on the $3.12^2$ lattice given in \cite{Jensen1998}, as might
be expected. In fact, Hammersley \cite{H61} proved that the growth
constants for SAWs and SAPs are equal on the $d$-dimensional
hypercubic lattice, for any given lattice dimensionality $d.$ The
proof of this result depends on a simple geometric unfolding argument
that becomes extremely messy on the $3.12^2$ lattice. As a result, it
is unlikely that one would wish to prove equality in that way. The
explicit mapping we have given is therefore a more appropriate route
to the proof.

\section{Covering lattice results}
\label{sec covering}

A result of Watson \cite{Watson1970} gives a bijection between SAWs on
a given lattice $\mathcal{L}$ and \emph{neighbour avoiding walks}
(NAW) on a related lattice called the \emph{covering lattice}
$\mathcal{L}^c$. This bijection means that we can
also determine the number of NAWs on the covering lattice of the
$3.12^2$ lattice.

First let us define the above terms. A neighbour avoiding walk is a
SAW with the additional restriction that adjacent occupied vertices 
must be connected by a single edge of the walk.  Given a graph
$\mathcal{L}$ its covering graph $\mathcal{L}^c$ is defined as
follows:
  \begin{itemize}
  \item For each edge $e_{i}$ in the edge set of $\mathcal{L}$  create
    a vertex in $\mathcal{L}^c$.
  \item If two edges $e_{i}, e_{j}$ in the edge set of $\mathcal{L}$ meet at the same
    vertex then create an edge in $\mathcal{L}^c$ connecting the
    vertices corresponding to $e_{i}$ and $e_j$ in $\mathcal{L}^c$.
  \end{itemize}
  
\begin{figure}[h!]
  \centering
  \includegraphics[scale=0.5]{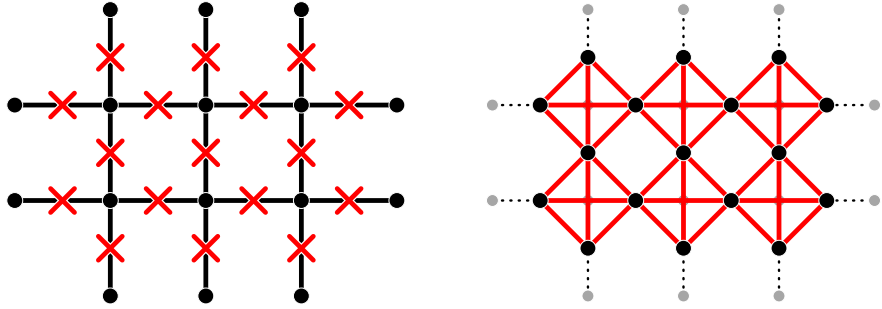}
  \caption{(left) A portion of the square lattice with a vertex placed at each edge
    (marked with a cross) and (right) the corresponding portion of the
    covering lattice of the square lattice --- each vertex of the
    covering lattice corresponds to an edge in the square lattice, and
    two vertices are connected by an edge if the corresponding edges
    in the square lattice meet at a vertex.}
  \label{fig sqr cover}
\end{figure}

\begin{figure}[h!]
  \centering
  \includegraphics[width=150mm, height=65mm]{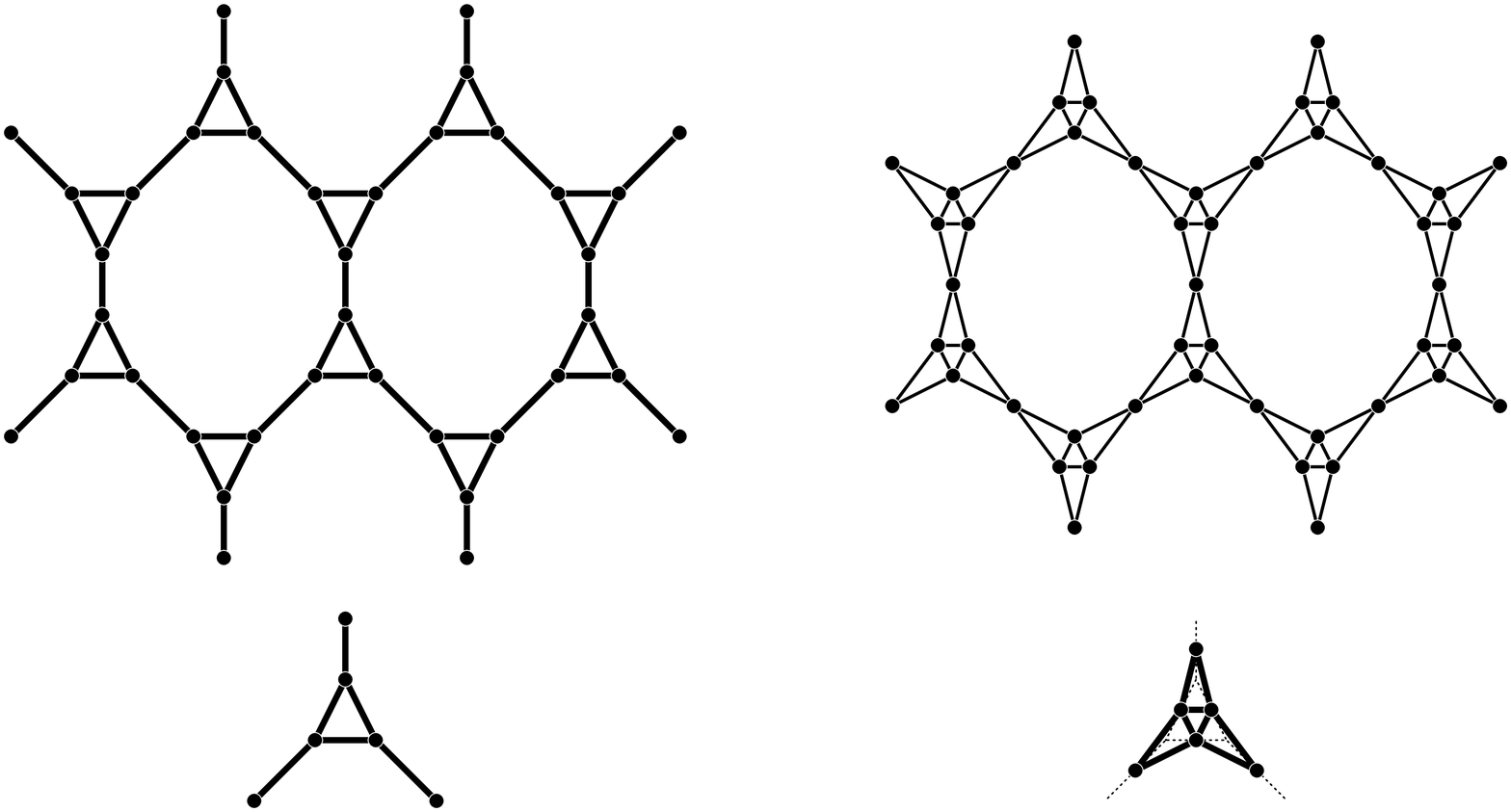}
  \caption{A portion of the $3.12^2$ lattice and the corresponding
    portion of its covering lattice. Below each lattice we show a unit
    cell of the $3.12^2$ lattice and the corresponding cell in the
    covering lattice.}
  \label{fig 312 cover}
\end{figure}

  In Figure~\ref{fig sqr cover} we show the covering lattice of the
  square grid, in Figure~\ref{fig 312 cover} the covering lattice of
  the $3.12^2$ lattice and in Figure~\ref{fig hex kag} the covering
  lattice of the hexagonal lattice. Note that the covering lattices of
  the $3.12^2$ and hexagonal lattices are tetra-valent.

\begin{theorem}[\cite{Watson1970}]
  There is a bijection between $n$-edge self-avoiding walks
  on the lattice $\mathcal{L}$ and $n-1$-edge neighbour avoiding walks
  on the covering lattice $\mathcal{L}^c$.
\end{theorem}
\proof 
A self-avoiding walk of $n$-edges may be encoded as a list of edges,
$[e_1, e_2, \dots, e_n]$, such that $e_i \neq e_j$ (for $i\neq j$) and
$e_i$ and $e_j$ meet at a vertex if and only if $i = j \pm 1$.
Mapping these edges to the corresponding vertices on the covering
lattice we obtain a list of vertices $[v_1, v_2, \dots, v_n]$ such
that $v_i \neq v_j$ and such that $v_i$ and $v_j$ are joined by an
edge if an only if $i = j \pm 1$. This is just an $n-1$ edge
neighbour-avoiding walk on the covering lattice. 
\qed

\begin{cor}
  The growth constant for neighbour avoiding walks on the covering
  lattice of the $3.12^2$ lattice is equal to that of SATs and SAWs on
  the $3.12^2$ lattice and is:
  \begin{displaymath}
    \mu = 1.711041297\dots
  \end{displaymath}
\end{cor}

\subsection{Mapping onto SAWs and into SATs.}
For any tri-valent lattice $\mathcal{L}$, we can define a mapping from the
set of SAWs, to a set $V$ of walks on the covering lattice
$\mathcal{L}^c$.

\begin{figure}[h!]
  \centering 
  \includegraphics[scale=0.35]{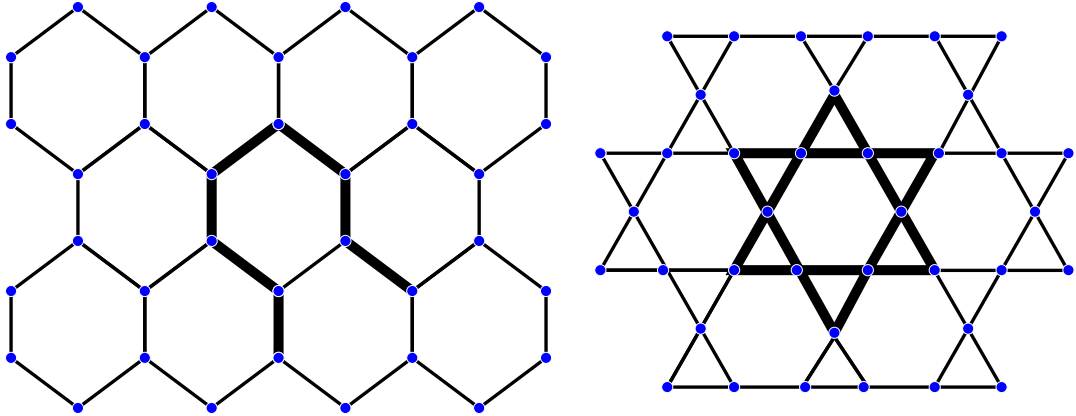}
  \caption{(left) A self-avoiding walk on the hexagonal lattice is 
    shown with bold lines. (right)~The image set of this walk on
    the covering $3.6.3.6$ lattice is shown with bold lines. }
  \label{fig hex kag}
\end{figure}

For a given self-avoiding walk on $\mathcal{L}$, with edge-sequence
$\{e_1, e_2, \dots, e_n\}$, map each edge $e_i$, except the last, to
two walks on the covering lattice $\mathcal{L}^c$, one being a single edge
of a triangle, the second being the other two edges. Thus one is of length 1:
the edge in $\mathcal {L}^c$ that connects $e_i$ and $e_{i+1}$, and the other is of
length 2: the 2-stepped self-avoiding walk in $\mathcal{L}^c$ that
connects $e_i$ and $e_{i+1}$. The image set
$V=V(\mathcal{L}^c)$ of the set of self-avoiding walks on $\mathcal{L}$
under this mapping is a set of walks on $\mathcal{L}^c$. An example of
the image set of a self-avoiding walk under this mapping is shown in
Figure \ref{fig hex kag}.

Let $W$ and $T$ denote the set of, respectively, SAWs and SATs on
$\mathcal{L}^c$. To make all walks in the image set start at the same
vertex, corresponding to the edge $e$ in $\mathcal{L}$, assume that all 
self-avoiding walks on $\mathcal{L}$ start with the edge $e.$
In particular, all have at least one edge.

\begin{lemma}
It follows that $W\subseteq V \subseteq T.$  
\end{lemma}
\begin{proof}
  We first show that the mapping is into SAT. Each subwalk $e_i, v_i,
  e_{i+1}$ of a self-avoiding walk $w$ on $\mathcal{L}$ uniquely determines a
  triangle, corresponding to the vertex $v_i$, and one or two edges of
  that triangle that appear in the image walk. As all vertices in $w$
  are distinct, no edge in $\mathcal{L}^c$ can appear twice in the image
  walk.
  
  For the other inclusion, consider a self-avoiding walk $\tilde
  w=\{e_1, e_2, \dots , e_n\}$ on $\mathcal{L}^c$.  Decompose $\tilde w$
  into subwalks by the following rule.  Set $i=1$ and $k=1$.

  Let $W_k=\{e_i, e_{i+1}\}$ if $e_i$ and $e_{i+1}$ are parts of different
  triangles, and let $i=i+2$ and $k=k+1$. Otherwise let $W_k=\{e_i\}$, and
  let $i=i+1$ and $k=k+1$. Repeat.
  
  Each subwalk $W_k$ corresponds to a unique triangle in $\mathcal{L}^c$,
  and so to a unique vertex $v_k$ in $\mathcal{L}$. It follows from the
  construction that $v_k$ is adjacent to $v_{k-1}$ and $v_{k+1}$ (if
  applicable). Let $w$ be the walk with vertex sequence $\{v_1, v_2,
  \dots v_K\}$, where $K$ is the index of the last subwalk.  Since at
  most two edges (which must then be adjacent in the walk) of a
  SAW $\tilde w$ can be part of any given triangle, the
  walk $w$ is self-avoiding. 
\end{proof}

Letting $w_n$ denote the number of different
walks of length $n$ in $V$, we may thus define a growth constant
$\omega=\lim_{n\to\infty}w_n^{1/n}$, and a generating function 
$V(z)=\sum_nw_nz^n$, with radius of convergence $z_c=1/\omega$.
By the lemma, it follows immediately that
\begin{displaymath}
  \mu_{W}(\mathcal{L}^c) \leq \omega(\mathcal{L})\leq\mu_{T}(\mathcal{L}^c),
\end{displaymath} 
where $\mu_{W}$ and $\mu_{T}$ are the growth constants
for SAWs and SATs, respectively.
Let $L(z)$ be the generating function for SAWs on $\mathcal{L}$ (with
the above restriction on the first step).  Since
$(z+z^2)V(z)=L(z+z^2)$, we have $\mu^{-1}=\omega_{W}(\mathcal{L})^{-1}
+\omega_{W}(\mathcal{L})^{-2}$.
An immediate consequence of this equation is the following theorem:

\begin{theorem} Let $\mu=\mu_{W}(\mathcal{L})$. 
  \begin{displaymath} 
    \omega=\frac12\left(\mu+\sqrt{4\mu+\mu^2}\right).
  \end{displaymath} 
\end{theorem}
  
\subsubsection{The hexagonal lattice}
It is well known that the covering lattice of the hexagonal lattice is
the Kagom\'e, or $3.6.3.6$, lattice. Using Nienhuis' value
$\mu_{W}(6^3)=\mu=\sqrt{2+\sqrt{2}}$ we find

\begin{displaymath}
  \mu_{W}(3.6.3.6)\leq\omega(3.6.3.6)= 2.5674465\ldots \leq\mu_{SAT}(3.6.3.6).
\end{displaymath}
Using the rigorous bounds $1.841925 <\mu_{W}(6^3)<1.868832$, 
due to \cite{J04} and \cite{Alm2004} respectively, we get
\begin{displaymath}
  \mu_{W}(3.6.3.6)\leq\omega(3.6.3.6)<2.5903041,
\end{displaymath}
and
\begin{displaymath}
  2.561114 <\omega(3.6.3.6)\leq\mu_{T}(3.6.3.6).
\end{displaymath}

These values should be compared with the estimate
$\mu_{W}(3.6.3.6)\approx 2.560576765$, from \cite{J04}, and the previous best
upper bound $\mu_{W}(3.6.3.6)\leq 2.60493$, from \cite{Alm2004b}.

\subsubsection{The $3.12^2$ lattice}
Using the Nienhuis value, we get bounds for the covering lattice
$(3.12^2)^c$ as follows
\begin{displaymath}
  \mu_{W}((3.12^2)^c)\leq 2.4185167\ldots \leq \mu_T((3.12^2)^c),
\end{displaymath}
and using the rigorous bounds given in \cite{J04} and \cite{Alm2004}, 
\begin{displaymath}
  \mu_{W}((3.12^2)^c)<2.4274958 \mbox{, and } 2.4157954 <\mu_T((3.12^2)^c).
\end{displaymath}

\begin{rem}
  An analogous mapping may be defined for lattices with higher
  coordination number. However, as the co-ordination number increases, the difference
  between the SAW and SAT growth constants increases.
\end{rem}

\begin{rem}
  In a similar way, an improved upper bound for the dual $D(3.12^2)$
  of the $3.12^2$ lattice may be derived. 
  Note that the triangular lattice $3^6$ is a subgraph of
  $D(3.12^2)$. Map each edge $e$ in a SAW on $3^6$ to three walks on
  $D(3.12^2)$: the edge $e$ itself, and the two walks of length 2
  that connect the endpoints of the edge $e$. The image set $I$ of
  the mapping contains all self-avoiding walks on $D(3.12^2)$. The
  growth constants for these two sets are related by
  $\mu_{W}(3^6)^{-1}=\mu_{I}(D(3.12^2))^{-1}+2\mu_{I}(D(3.12^2))^{-2}$.
  Using the upper bound $\mu_{W}(3^6)<4.25142$, \cite{Alm2004b}, we get
  $\mu_{W}(D(3.12^2))< 5.73424$, a slight improvement of the
  previous upper bound, 5.79621, \cite{Alm2004b}.
\end{rem}

\section{Conclusions}
We have found an expression for the number of self-avoiding walks and
self-avoiding trails on the $3.12^2$ lattice in terms of the
generating functions of SAWs, returns and tadpoles on the hexagonal
lattice. 

A bijection allows us to then express number of neighbour avoiding
walks on the covering lattice of the $3.12^2$ lattice in terms of
these same objects on the hexagonal lattice. Since the growth constant
of walks is known (non-rigorously) on the hexagonal lattice, our
expressions also give the growth constant for SAWs and SATs on the
$3.12^2$ lattice and NAWs on the covering lattice. This last lattice
is tetra-valent.

A similar mapping, which is {\em into} the set of SATs and {\em onto} the set of
SAWs on the covering lattice, gives bounds for the growth constants on
the $3.6.3.6$ lattice and the covering lattice of $3.12^2$ lattice.

\section*{Acknowledgements}
All authors are supported by the Australian Research Council, and wish to express their
gratitude for that support.

\bibliographystyle{plain}
\bibliography{saw312_bib.bib}

\begin{table}[h!]
  \centering
  \begin{tabular}{|c|| r | r |}
    \hline
    $n$ & SAW & Trails \\
    \hline
    0 & 1 & 1 \\
    1 & 3 & 3 \\
    2 & 6 & 6 \\
    3 & 10 & 12 \\
    4 & 18 & 22 \\
    5 & 32 & 40 \\
    6 & 56 & 72 \\
    7 & 100 & 128 \\
    8 & 176 & 224 \\
    9 & 312 & 400 \\
    10 & 552 & 704 \\
    11 & 976 & 1248 \\
    12 & 1724 & 2208 \\
    13 & 3018 & 3900 \\
    14 & 5240 & 6870 \\
    15 & 9078 & 12002 \\
    16 & 15780 & 20860 \\
    17 & 27502 & 36232 \\
    18 & 47952 & 63086 \\
    19 & 83602 & 110002 \\
    20 & 145700 & 191808 \\
    21 & 253666 & 334388 \\
    22 & 440696 & 582590 \\
    23 & 763624 & 1013674 \\
    24 & 1321176 & 1760024 \\
    25 & 2286260 & 3049440 \\
    26 & 3959928 & 5278204 \\
%    27 & 6861692 & 9138436 \\
%    28 & 11886772 & 15832512 \\
%    29 & 20581946 & 27436156 \\
%    30 & 35619908 & 47529862\\
    \hline
  \end{tabular}
  \caption{The first few terms of the generating functions for SAWs
    and SATs on the $3.12^2$ lattice. The data was generated using a
    backtracking algorithm.}
  \label{tab 312 series}
\end{table}

\begin{table}[h!]
  \centering
  \begin{tabular}{|c|| r | r | r | r | r |}
    \hline
    $n$ & SAW & Returns & Tadpoles & Dumbbells & Theta graphs \\
    \hline
    1  &  3  &  0  &  0  &  0  &  0  \\
    2  &  6  &  0  &  0  &  0  &  0  \\
    3  &  12  &  0  &  0  &  0  &  0  \\
    4  &  24  &  0  &  0  &  0  &  0  \\
    5  &  48  &  0  &  0  &  0  &  0  \\
    6  &  90  &  6  &  0  &  0  &  0  \\
    7  &  174  &  0  &  3  &  0  &  0  \\
    8  &  336  &  0  &  6  &  0  &  0  \\
    9  &  648  &  0  &  12  &  0  &  0  \\
    10  &  1218  &  30  &  24  &  0  &  0  \\
    11  &  2328  &  0  &  54  &  0  &  $1\frac{1}{2}$  \\
    12  &  4416  &  24  &  108  &  0  &  0  \\
    13  &  8388  &  0  &  222  &  $1 \frac{1}{2}$  &  0  \\
    14  &  15780  &  168  &  414  &  3  &  3  \\
    15  &  29892  &  0  &  834  &  6  &  9  \\
    16  &  56268  &  288  &  1614  &  9  &  3  \\
    17  &  106200  &  0  &  3168  &  33  &  12  \\
    18  &  199350  &  1170  &  5940  &  63  &  24  \\
    19  &  375504  &  0  &  11598  &  120  &  $64\frac{1}{2}$  \\
    20  &  704304  &  2760  &  21972  &  225  &  54  \\
    21  &  1323996  &  0  &  42306  &  534  &  147  \\
    22  &  2479692  &  9504  &  79398  &  975  &  219  \\
    23  &  4654464  &  0  &  152460  &  1953  &  546  \\
    24  &  8710212  &  25776  &  286470  &  3672  &  627  \\
    25  &  16328220  &  0  &  546102  &  $7627\frac{1}{2}$  &  1536  \\
    26  &  30526374  &  84006  &  1023030  &  14103  &  2127  \\
    \hline
  \end{tabular}
  \caption{The first few terms of the generating functions of SAWs,
    returns, tadpoles, dumbbells and theta graphs on the hexagonal
    lattice. The data was generated by a backtracking algorithm and from data supplied by M. F. Sykes.}
  \label{tab hex series}
\end{table}

\end{document}